\documentclass[12pt,a4paper]{article}
\usepackage[latin1]{inputenc}
\usepackage{amsmath}
\usepackage{amsfonts}
\usepackage{amssymb}
\usepackage{makeidx}

\begin{document}
\sloppy

\title{\textbf{Irreversibility and maximum generation in $\kappa$-generalized statistical mechanics}}
\author{Umberto Lucia\\I.T.I.S. ``A. Volta'', Spalto Marengo 42, 15121 Alessandria, Italy}
\date{}
\maketitle

\begin{abstract}
This paper develops an analytical relation on probability to obtain a condition on stability of steady states of open systems which  represents the condition of quasi-ergodicity for the open systems too. It is proved that the result obtained is consequence of the maximum entropy generation principle.

\end{abstract}

\section{Introduction}
In phenomena out of equilibrium irreversibility manifests itself because the fluctuations of the physical quantities, which bring the system apparently out of stationarity, occur symmetrically about their average values (Gallavotti 2006). The $\varepsilon$-stady state definition allows us to obtain that for certain fluctuations the probability of occurrence follows a universal law and the frequency of occurrence is controlled by a quantity that has been related to the entropy generation. Moreover, this last quantity has a purely mechanical interpretation which is related to the the ergodic hypothesis which proposed that an isolated system evolves in time visiting all possible microscopic states. Moreover, considering that the open system is a system with perfect accessibility represented as a probability space in which is defined a $PA$-measure and a Borel function process the ergodic hypothesis itself is a consequence of the $\varepsilon$-steady state definition because of the Hypothesis \ref{regular process}, introduced in Section 4.

The principle of maximum entropy generation represents the macroscopic effect of these theories, which, conversely, are its statistical interpretation. As a consequence of this principle it has been proved the irreversible ergodicity (Lucia 2008). As H. Primas has pointed out (Primas 1999): `The importance of the ergodic theorem lies in the fact that in most applications we see only a single individual trajectory that is a particular realization of the thermostatic process. The Kolmogorov's theory of stochastic processes refers to equivalence classes of functions and the Birkoff's ergodic theorem provides a crucial link between the ensamble description and the individual description of chaotic phenomena.'. The modern concept of subjective probabilities implies a behaviour based on the Boolean logic (Primas 1999). The result obtained represents also the relation between the probability measurement in the probability space and real statistical facts. Following H. Primas (Primas 1999) the set-theoretical representation of the Boolean algebra in terms of a Kolmogorov probability space is useful because it allows to relate a dynamic in terms of a probabilistic density and all known context-independent physical laws are formulated in terms of pure states (Primas 1999).

By entropy generation it has been introduced the global thermodynamic effect of the action of the forces finding a global expression which links the microscopical values to the macroscopical thermodynamic quantities (Lucia 2008). On the other hand, it has been argued that the information of a complex system may be incomplete because the dynamics of the system can be only partially known consequently to its not complete accessibility (Wang 2004). In addition the $PA$-measure in the probability space is not yet univocally defined, and it is difficulty to develop a statistical thermodynamics for an irreversible system. In 2009 the integral expression of the $PA$-measure in the probability space, useful in statistical thermodynamics of the complex and irreversible systems, has been obtained (Lucia 2009). To do it the basis of the incomplete information on complex system was developed, underlying that in non-equilibrium transformation the volume of the phase space contracts; consequently, entropy generation were related to the incomplete information theory by using the Wang development of the Tsallis theory. This result underlines that the relevant physical quantities for the stochastic analysis of the irreversibility are the probability of the state in the phase space, the Hamiltonian valued at the end points of trajectory, and time (Lucia 2009). 

But, the Hamiltonian valued at the end points of trajectory in the phase space is not easy to be evaluated. In recent years, the study of an increasing numbers of natural phenomena that appear to deviate from standard statistical distributions has kindled interest in alternative formulation of statistical mechanics (Kaniadakis 2005). The new formulations of statistical mechanics should preserve most of the mathematical and epistemological structure of the Boltzmann-Gibbs theory, while reproducing the phenomenology of the anomalous systems (Kaniadakis 2005). This $\kappa$-theory obtained a lot of results in the analysis of natural phenomena, but it was not extended to irreversible open systems. A modified $\kappa$-theory has been proposed to use it in irreversible open systems, in order to obtain the integral value of the Hamiltonian at the end points of trajectory in the phase space, obtaining a new statistical expressions for the entropy generation (Lucia 2010). 

The aim of this paper is to find an analytical relation on probability to obtain a condition on stability of steady states of open systems which  represents the condition of quasi-ergodicity for the open systems too.

To do so in Section 2 the open system is defined, in Section 3 the entropy generation is introduced and related to its definition in rational thermodynamics, in Section 4 the quasi-ergodicity is discussed in relation to the statistical properties of entropy generation whose definition related to $\kappa-$statistics is introduced in Section 5 where the required results is obtained.

\section{The dynamical system}
In this section it will be defined the system considered. To do so, we must consider the definition of `system with perfect accessibility', which allows us to define both the thermodynamic system and the complex and dynamical system.

\newtheorem{definition}{Definition}
\newtheorem{comment}{Comment}

\begin{definition} \emph{(Lucia 2008)}
 - A system with perfect accessibility $\Omega_{PA}$ is a pair $\left(\Omega, \Pi\right)$, with $\Omega :=\{\mathbf{\sigma}_i\in\mathbb{R}^{6N}:\mathbf{\sigma}_i=(\mathbf{p}_i,\mathbf{q}_i), i\in[1,N]\}$, being $N$ the number of particles, $\mathbf{q}_i\in\mathbb{R}^{6N}$, $i\in[1,N]$ their canonical coordinates and $\mathbf{p}_i\in\mathbb{R}^{6N}$, $i\in[1,N]$ their conjugate momenta, and $\Pi$ a set whose elements  $\pi$ are called process generators, together with two functions:
\begin{equation}
\pi \mapsto \mathcal{S}
\end{equation}
\begin{equation}
\left(\pi^{'},\pi{''}\right) \mapsto \pi^{''}\pi{'}
\end{equation}
where $\mathcal{S}$ is the state transformation induced by $\pi$, whose domain $\mathcal{D}\left(\pi\right)$ and range $\mathcal{R}\left(\pi\right)$ are non-empty subset of $\Omega$.
This assignment of transformation to process generators is required to satisfy the following conditions of accessibility:
\begin{enumerate}
	\item $\Pi\sigma:=\left\{\mathcal{S}\sigma :\pi\in\Pi,\sigma\in \mathcal{D}\left(\pi\right)\right\}=\Omega$ , $\forall \sigma\in\Omega$\emph{:} the set $\Pi\sigma$ is called the set of the states accessible from $\sigma$ and, consequently, it is the entire \emph{state space}, the phase spase $\Omega$;
	\item if $\mathcal{D}\left(\pi ''\right)\cap \mathcal{R}\left(\pi '\right)\neq \emptyset\Rightarrow \mathcal{D}\left(\pi ''\pi '\right)=\mathcal{S}_{\pi '}^{-1}\left(\mathcal{D}\left(\pi ''\right)\right)$ and $\mathcal{S}_{\pi ''\pi '}\sigma =\mathcal{S}_{\pi ''}\mathcal{S}_{\pi '}\sigma, \forall\mathbf{\sigma}\in \mathcal{D}\left(\pi ''\pi '\right)$
\end{enumerate}
\end{definition}

\begin{definition} \emph{(Lucia 2001)} 
- A process in $\Omega_{PA}$ is a pair $\left(\pi ,\sigma\right)$, with $\sigma$ a state and $\pi$ a process generator such that $\sigma$ is in $\mathcal{D}\left(\pi\right)$. The set of all processes of $\Omega_{PA}$ is denoted by:
\begin{equation}
\Pi \diamond\Omega =\left\{\left(\pi ,\sigma\right): \pi \in \Pi ,\sigma\in \mathcal{D}\left(\pi\right)\right\}
\end{equation}
If $\left(\pi ,\sigma\right)$ is a process, then $\sigma$ is called the initial state for $\left(\pi ,\sigma\right)$ and $\mathcal{S}\sigma$ is called the final state for $\left(\pi ,\sigma\right)$.
\end{definition}

\begin{definition} \emph{(Lucia 2001)} 
\label{thermdef} - A thermodynamic system is a system with perfect accessibility $\Omega_{PA}$ with two actions $W\left(\pi ,\sigma\right)\in\mathbb{R}$ and $H\left(\pi ,\sigma\right)\in\mathbb{R}$, called work done and heat gained by the system during the process $\left(\pi ,\sigma\right)$, respectively.
\end{definition}

\newtheorem{hypothesis}{Hypothesis}
\begin{hypothesis} \emph{(Lucia 2008)} 
- There exists a statistics $\mu _{PA}$ describing the asymptotic behaviour of almost all initial data in perfect accessibility phase space $\Omega_{PA}$ such that, except for a volume zero set of initial data $\mathbf{\sigma}$, it will be:
\begin{equation}
\lim_{T\longrightarrow\infty}\frac{1}{T}\sum^{T-1}_{j=1}\varphi\left(\mathcal{S}^{j}\mathbf{\sigma}\right)=\int_{\Omega}\mu_{PA}\left(d\mathbf{\sigma}\right)\varphi\left(\mathbf{\sigma}\right)
\end{equation}
for all continuous functions $\varphi$ on $\Omega_{PA}$ and for every transformation $\mathbf{\sigma}\mapsto \mathcal{S}_{t}\left(\mathbf{\sigma}\right)$.
\end{hypothesis}

\begin{definition}
 - The triple $\left(\Omega_{PA},\mathcal{F},\mu_{PA}\right)$, with $\mathcal{F}$ an algebra or a $\sigma-$algebra, is a \emph{measure space}, the \emph{Kolmogorov probability space} $\Gamma$.
\end{definition}

\begin{definition}
 - A dynamical law $\tau$ is a group of meausure-preserving automorphisms $\mathcal{S}:\Omega_{PA}\rightarrow\Omega_{PA}$ of the probability space $\Gamma$.
\end{definition}

\begin{definition} \emph{(Lucia 2008)}
 - A dynamical system $\Gamma_{d}=\left(\Omega_{PA},\mathcal{F},\mu_{PA},\tau\right)$ consists of a dynamical law $\tau$ on the probability space $\Gamma$.
\end{definition}

\section{Entropy generation}
In thermodynamics, the energy lost for irreversible processes is evaluated by the first and second law of thermodynamics for the open systems. So the following definition can be introduced:
\begin{definition} \emph{(Lucia 2009)}
 - The entropy generation $S_{g}$ is defined as:
\begin{equation}
\frac{W_{lost}}{T_{ref}}=S_{g}:=\frac{Q_{r}}{T_{a}}\left( 1- \frac{T_{a}}{T_{r}} \right)+\frac{\Delta H}{T_{a}}-\Delta S+\frac{\Delta E_{k}+\Delta E_{g}-W}{T_{a}}
\end{equation}
where $W_{lost}$ is the work lost for irreversibility, $T_{ref}$ the temperature of the lower reservoir, $Q_{r}$ is the heat source, $T_{r}$ its temperature, $T_{a}$ is the ambient temperature, $H$ is the enthalpy, $S$ is the entropy, $E_{k}$ is the kinetic energy, $E_{g}$ is the gravitational one and $W$ is the work.
\end{definition}

Following Truesdell (Truesdell 1970), for each continuum thermodynamic system (isolated or closed or open), in which it is possible to identify a thermodynamics subsystem with elementary mass $dm$ and elementary volume $dV=dm/\rho$, with $\rho$ mass density (Truesdell 1970, Lucia 1995, Lucia 2001, Lucia 2008) the thermodynamic description can be developed by referring to the generalized coordinates $\left\{\xi_{i},\dot{\xi}_{i},t\right\}_{i\in[1,N]}$, with $\xi_{i} =\alpha_{i}-\alpha^{\left(0\right)}_{i}$, $\alpha_{i}$ the extensive thermodynamic quantities and $\alpha^{\left(0\right)}_{i}$ their values at the stable states.

\newtheorem{theorem}{Theorem}
\begin{theorem} \emph{(Lucia 1995)}
- The termodynamic Lagrangian can be obtained as:
\begin{equation}
	\mathcal{L}=-T_{ref}\,S_{g}
		\label{Lagrangian}
\end{equation}
\end{theorem}

\begin{theorem}
- \label{maxentrth} \label{maxentrp} \textbf{The principle of maximum entropy generation} \emph{(Lucia 1995, Lucia 2008):}
 The condition of stability for the open system' stationary states is that its entropy generation $S_g$ reaches its maximum:
\end{theorem}

\section{Irreversible ergodicity}
A real-valued random variable $\pi :\Omega_{PA}\rightarrow \mathbb{R}$ on $\Gamma$ induces a probability measurement $\mu_{PA}:\mathcal{F}\rightarrow\left[0,1\right]$ on the state space $\left(\mathcal{F}_{\mathbb{R}},\mathbb{R}\right)$.
Considering the probability as a property of the generating conditions of a sequence, randomness can be related to predictability and retrodictability \emph{(Primas 1999)}. 
A family $\left\{\mathbb{\xi}\left(t\right):t\in\mathcal{R}\right\}$ is called a stocastic process, which can be represented by a family $\left\{\gamma\left(\mathbf{\sigma}\left(t\right)\right):t\in\mathbb{R}\right\}$ of equivalent classes of random variables $\mathbf{\xi}\left(t\right)$ on $\Gamma$. The point function $\gamma\left(\mathbf{\sigma}\left(t\right)\right)$ is called trajectory of the stocastic process $\mathbf{\xi}\left(t\right)$. 
The description of physical systems in terms of a trajectory of a stochastic process corresponds to a point dynamics, while its description in terms of equivalent classes of trajectories and an associated probability measure corresponds to an ensemble dynamics \emph{(Primas 1999)}.

\begin{definition}
 - A stochastic process is said weakly stationary if:
\begin{enumerate}
	\item $\left[\xi^{2}\left(t\right)\right]_{ev}, \forall t\in\mathbb{R}$
	\item $\xi_{ev}\left(t+\tau\right)=\xi_{ev}\left(t\right), \forall t,\tau\in\mathbb{R}$
	\item $\left[\xi\left(t_{\alpha}+\tau\right)\xi\left(t_{\beta}+\tau\right)\right]_{ev}=\left[\xi\left(t_{\alpha}\right)\xi\left(t_{\beta}\right)\right]_{ev}, \forall  t_{\alpha},t_{\beta},\tau\in\mathbb{R}$.
\end{enumerate}
\end{definition}

A consequence of the Wiener-Khintchin-Einstein theorem (Primas 1999) consists in the existence of a complex-valued function $\xi:\mathbb{R}\rightarrow\mathbb{C}$, continuous at the origin, which is the covariance function of a complex-valued second-order, weakly stationary and continuous stochastic process if and only if it can be represented as:
\begin{equation}
\xi\left(t\right)=\int_{-\infty}^{\infty}e^{i\lambda t}d\hat{\xi}\left(\lambda\right)
\end{equation}
where, as a consequence of the Bochner-Cramr representation theorem \emph{(Primas 1999)}, $\hat{\xi}:\mathbb{R}\rightarrow\mathbb{R}$ is a real, never decreasing and bounded function, called spectral distribution function of the stochastic process. Moreover, the Lebesgue's decomposition theorem states that the spectral distribution function can be decomposed uniquely as:
\begin{equation}
\hat{\xi}=c^{d}\hat{\xi}^{d}+c^{s}\hat{\xi}^{s}+c^{ac}\hat{\xi}^{ac}
\end{equation}
with $c^{d}\geq 0$, $c^{s}\geq 0$, $c^{ac}\geq 0$, $c^{d}+c^{s}+c^{ac}=1$, $\hat{\xi}^{d}$ step function, $\hat{\xi}^{s}$ continuous and singular real function, $\hat{\xi}^{ac}$ absolutely continuous real function.

There exists a close relationship between regular and stochastic processes and the irreversibility. A system is called irreversible if the lost energy is strictly positive. Following K\"onig and Tobergte (Primas 1999) a linear input-output system behaves irreversible if and only if the associated distribution function fulfills the Wiener-Krein criterion for the spectral density of a linear regular stochastic process: a weakly stationary stochastic process $\xi$ with mean value $\xi_{ev}=0$ and spectral distribution $\lambda\mapsto\hat{\xi}\left(\lambda\right)$ is linearly regular if and only if its spectral distribution function is absolutely continuous and if:
\begin{equation}
\int_{-\infty}^{\infty}\frac{\ln \left(d\hat{\xi}\left(\lambda\right)/d\lambda\right)}{1+\lambda^{2}}d\lambda >-\infty
\end{equation}

Following Wiener and Akutowics, we can assume the following:
\begin{hypothesis} \label{regular process} \emph{(Primas 1999)}
 Every stationary process with absolutely continuous spectral function \emph{(=} regular process\emph{)} is ergodic.
\end{hypothesis}

The mathematical foundation of ergodic theory is to establish a connection between phase average and time average (van Lith 2001). It follows an association between ergodic theory and objective interpretation of probability because the ergodic theory  establishes a connection between probability measures and objective features of real word (van Lith 2001). A probability distribution is \textit{stationary} if it is constant at all fixed points in $\Gamma$, and it reflects the fact that the system is in a \textit{steady} state. Modern ergodicity is founded on the concept of measure preserving dynamical systems $\Gamma_{d}$ and on the Birkhoff's ergodic theorem (van Lith 2001). Here, considering the previous Section, we introduce the stochastic processes. To do so, we state that:

\begin{theorem}
 - If $\mu_{PA}\left(\Omega_{PA}\right)$ is finite, then for any integrable function $\varphi :\Omega_{PA}\rightarrow \mathbb{R}$, the time average $\left\langle\varphi\right\rangle_{t}$ on $\gamma$ is defined for all orbits $\gamma$ outside of a set $\mathcal{N}_{\varphi}$ of measure $\mu\left(\mathcal{N}_{\varphi}\right)= 0$. Furthermore $\left\langle\varphi\right\rangle_{t}$ is integrable, with $\left\langle\varphi\right\rangle_{t}\circ \gamma = \left\langle\varphi\right\rangle_{t}$ wherever it is defined, and with
\begin{equation}
\int _{\Omega_{PA}}\left\langle\varphi\right\rangle_{t}d\Omega_{PA}=\int _{\Omega_{PA}}\varphi d\Omega_{PA}
\end{equation}
\end{theorem}

\newtheorem{corollary}{Corollary}
\begin{corollary} \label{birkoff}- \textbf{Birkoff's ergodic theorem}. A measure preserving transformation $\mathcal{S}:\Omega_{PA}\rightarrow\Omega_{PA}$ is ergodic if and only if, for every integrable function $\varphi:\Omega_{PA}\rightarrow\mathbb{R}$, the time average $\left\langle\varphi\right\rangle_{t}$ on $\gamma$ is equal to the space average $\int_{\Omega_{PA}}\varphi\left(\mathbf{\sigma}\right) \mu_{PA}\left(d\mathbf{\sigma}\right)$ for all points $\mathbf{\sigma}$ outside of some subset $\mathcal{N}_{\varphi}$ of measure $\mu_{PA}\left(\mathcal{N}_{\varphi}\right) = 0$.
\end{corollary}

Consequently, measurements results as the infinite time averages of phase functions because they take a long time compared to the microscopic relaxation time and, for metrically  transitive ($=$ ergodic) systems, measurement results are almost always equal to microcanonical averages (van Lith 2001). Moreover, the initial states lead to different paths in phase space, so the averages depend on the initial state.

\begin{definition} - \textbf{$\varepsilon$-steady state}.
 Let $\Gamma_{d}$ be a dynamical system and $\varepsilon=\left\{\varepsilon_{\mathcal{S}}\right\}$ be fixed and non zero. An open system is in $\varepsilon-$steady state during the time interval $\mathcal{T}$ if and only if for all $\mathcal{S}\in\Omega_{PA}$ there exists $\zeta_{\mathcal{S}}\in\mathbb{R}$ such that for all $t\in\mathcal{T}$ it follows:
\begin{equation}
\left|\left\langle \mathcal{S}\right\rangle_{\mathcal{T}}-\zeta_{\mathcal{S}}\right|\leq \varepsilon_{\mathcal{S}}
\end{equation}
\end{definition}

Consequently, the probability may fluctuate within small bounds and, consequently, dynamical evolution towards steady states is allowed. Every statistical observable is induced by a random variable, while an observable, that is a $\sigma$-homomorphism, defines only an equivalence class of random variables which induce this homomorphism. For a statistical description it is not necessary to know the point function $\sigma\mapsto\mathcal{S}\left(\sigma\right)$, but it is sufficient to know the observable $\mathbf{\xi}$. The description of a physical system in terms of an individual function $\pi:\sigma\mapsto\mathcal{S}\left(\sigma\right)$ distinguish between different points $\sigma\in\Omega_{PA}$ and corresponds to an individual description of equivalence classes of random variables does not distinguishes between different points and corresponds to a statistical ensemble description.

\begin{definition}
 - Let it be  $\pi :\Omega_{PA}\rightarrow \mathbb{R}$ a real-valued Borel function such that $\mathbf{\sigma}\mapsto\mathcal{S}\left(\mathbf{\sigma}\right)$ is integrable over $\Omega_{PA}$ with respect to $\mu_{PA}$, the expectation value $\pi_{ev}$ of $\mathcal{S}\left(\mathbf{\sigma}\right)$ with respect to $\mu_{PA}$ is:
\begin{equation}
\pi_{ev}:=\int_{\Omega}\mathcal{S}\left(\mathbf{\sigma\mathbf}\right)\mu_{PA}\left(d\mathbf{\sigma}\right)
\end{equation}
\end{definition}

As a consequence of the theorem (\ref{maxentrth}) and the relation (\ref{irrentrdef}), it follows that:
\begin{theorem} - \textbf{Irreversible ergodicity} \emph{(Lucia 2008)}
- In non equilibrium transformation the volume of the phase space $\Omega_{PA}$  contracts indefinitely.
\end{theorem}

\section{Entropy generation and its statistics}
If the system is ergodic, the time-averages over infinite times coincide with the Gibbs phase-averages, but there remains open the problem that nothing is known concerning the time-averages for large but finite times. It was shown that the use of time-averages amounts to introducing a measure in phase space, suitably defined by the dynamics of the system. One can imagine that, if one has to deal with a metastable state ergodic behaviour is granted only on a times-scale much larger than the available one, then the orbits could exhibit some strange features, such as a non integer fractal dimension on the observed time-scale, which may prevent the use of the Gibbs measure. It has been suggested that in some problems where metastable states show up, one has to replace the Gibbs measure by the Tsallis one. In the present paper we show that, if a system has dynamical time-averages compatible with a Tsallis ensemble, then, on a certain time-scale, the orbits have a definite non integer fractal dimension. We also show that the diffusion process of the orbits in phase space is, in some sense, slower than in the full chaotic case corresponding to Gibbs measure. In order to establish a correspondence between dynamics and Tsallis distribution one has to solve an analytical problem by an exact correspondence between dynamics and Tsallis distribution, given only for continuous-time dynamical systems and not for mappings. In the case of a mapping, such a correspondence is obtained by introducing a suitable limiting procedure (Carati 2008).

In 1988 Tsallis postulated the physical relevance of one-parameter generalization of the entropy (Tsallis 1988), introducing a generalization of standard thermodynamics and of Boltzmann-Gibbs statistical mechanics. In 2002, Kaniadakis proved some properties for the generalized logarithm, so that we can introduce the following:

\begin{definition} \emph{(Kaniadakis 2003, Pistone 2009)}
The generalized exponential $\exp_{\{\kappa\}}$ is defined as follows:
\begin{equation}
\exp_{\{\kappa\}}(\tau)=\exp\bigg(\int_{0}^{\tau }\frac{dt}{\sqrt{1+\kappa^{2}t^{2}}}\bigg)= \Bigg\{
\begin{array}{l l} 
\big(\kappa \tau + \sqrt{1+\kappa^{2}\tau^{2}}\big)^{1/\kappa} & \mbox{if $\kappa\not\neq 0$} \\ 
\exp(\tau) &  \mbox{if $\kappa= 0$}
\end{array}
\end{equation}
\end{definition}

\begin{definition} \label{log} \emph{(Kaniadakis 2003, Pistone 2009)}
The generalized logarithm $\ln_{\{\kappa\}}$ is defined as follows:
\begin{equation}
\ln_{\{\kappa\}}(\tau)=\frac{\tau^\kappa - \tau^{-\kappa}}{2\,\kappa}
\end{equation}
\end{definition}

\begin{definition} \label{entr} \emph{(Lucia 2010)}
The statistical expression, for the irreversible-entropy variation, results:
\begin{equation}
S_{g}=-\sum_{\gamma}p_{\gamma}\ln_{\{\kappa\}}(p_{\gamma})
\label{irrentrdef}
\end{equation}
with
\begin{equation}
p_{\gamma}=\alpha \exp_{\{\kappa\}}\bigg(-\frac{E_{\gamma}-\mu}{\lambda T }\bigg)
\label{prob}
\end{equation}
$p_\gamma$ is the probability of the path $\gamma$, and $\alpha = \big[\sum_\gamma \exp\big(-\frac{1}{2}{\int_V \frac{H_{\gamma(0)}+H_{\Gamma (\tau)}}{k_B T}dV+\frac{\tau\sigma_\gamma}{2k_B}}\big)\big]^{-1}$, with $H=u-\sum_i \mu_i\rho_i$ the non-equilibrium generalisation of the grand-canonical Hamiltonian, $H_{\gamma(0)}$ and $H_{\gamma(\tau)}$ the values of $H$ at the end points of trajectory $\gamma$ in $\tau =0$ and $\tau$, and $\sigma_\gamma$ the time-averaged rate entropy production of $\gamma$, 
and $\lambda = 2 k_{B}$ are two arbitrary, real and positive parameters, $\mu=\beta_{1}/\beta_{2}$, with $\beta_{1}$ and $\beta_{2}$ Lagrange multipliers in the entropic functional
\begin{equation}
\mathcal{F} = S_{\kappa} - \beta_{1} \Bigg(\sum_{\gamma}p_{\gamma}-1\Bigg) - \beta_{2} \Bigg(\sum_{\gamma}p_{\gamma} E_{\gamma}-U\Bigg)
\end{equation}
which must be stationary for variations of $\beta_{1}$ and $\beta_{2}$, and where $E_{\gamma}$ is the energy of the $\gamma$-th path and $U$ is the total energy, $\ln_{\{\kappa\}}(x)$ is an arbitrary function, a generalization of the logarithm function, whose properties have been studied by Csiszar \emph{(Csiszar 1967)} and when $\ln_{\{\kappa\}}(x)=\ln(x)$ the generalized entropy reduces to the Boltzmann-Gibbs-Shannon entropy \emph{(Wada and Scarfone 2009)}. 
\end{definition}

\newtheorem{lemma}{Lemma}
\begin{lemma}
Let us consider the $\ln_{\kappa}p_\gamma$. Its first order derivative is:
\begin{equation}
\ln_{\kappa}'(p_\gamma)=\frac{d\ln_{\kappa}(p_\gamma)}{d(p_\gamma)}=\frac{\kappa}{p_\gamma}\,\ln_{\kappa}(p_\gamma)+p_\gamma^{-(\kappa +1)}
\end{equation}
\end{lemma}
\textit{Proof}.\\
From the Definition \ref{log} it follows that:
\begin{equation}
\begin{split}
\ln_{\kappa}'(p_\gamma)& =\frac{d\ln_{\kappa}(p_\gamma)}{d(p_\gamma)}=\frac{d}{dp_\gamma}\Bigg(\frac{p_\gamma^\kappa -p_\gamma^{-\kappa}}{2\,\kappa}\Bigg)=\\ & =\frac{\kappa\,p_\gamma^{\kappa -1} +\kappa\,p_\gamma^{-(\kappa+1)}}{2\,\kappa} = \frac{\kappa}{p_\gamma}\,\frac{p_\gamma^\kappa -p_\gamma^{-\kappa}+2\,p_\gamma^{-\kappa}}{2\,\kappa}=\\ &=\frac{\kappa}{p_\gamma}\,\frac{p_\gamma^\kappa -p_\gamma^{-\kappa}}{2\,\kappa}+\frac{\kappa}{p_\gamma}\,\frac{2\,p_\gamma^{-\kappa}}{2\kappa}=\\ & =\frac{\kappa}{p_\gamma}\,\ln_{\kappa}(p_\gamma)+p_\gamma^{-(\kappa +1)}
\end{split}
\end{equation}
\begin{flushright}
$\Box$
\end{flushright}

As a consequence of the theorem of maximum entropy generation the following inequality can be obtained the following condition:
\begin{theorem}
The condition of stability for the open system' stationary states is the following inequality for the probability of the paths:
\begin{equation}
\frac{-\sum_\gamma \ln_{\{\kappa\}}p_\gamma}{\sum_\gamma p_\gamma^{-\kappa}}\geq\frac{1}{1+\kappa}
\label{pr}
\end{equation}
\end{theorem}
\textit{Proof}.\\
From the Definition \ref{entr} it is possible to write:
\begin{equation}
\begin{split}
S_g ' &= \frac{d}{dp_\gamma}\Big(-\sum_{\gamma}p_{\gamma}\ln_{\{\kappa\}}(p_{\gamma})\Big)
= \\& = -\sum_\gamma \,\ln_{\kappa}(p_\gamma)-\sum_\gamma p_\gamma\,\Bigg(\frac{\kappa}{p_\gamma}\,\ln_{\kappa}(p_\gamma)+p_\gamma^{-(\kappa +1)}\Bigg) =\\ &=-\sum_\gamma \,\ln_{\kappa}(p_\gamma)-\sum_\gamma \kappa \,\ln_{\kappa}(p_\gamma)-\sum_\gamma p_\gamma^{-\kappa} = \\ & =-\sum_\gamma (1+\kappa) \,\ln_{\kappa}(p_\gamma)-\sum_\gamma p_\gamma^{-\kappa}
\end{split}
\end{equation}
and, considering the Theorem \ref{maxentrp}, it follows:
\begin{equation}
S_g '\geq 0 \Rightarrow -\sum_\gamma (1+\kappa) \,\ln_{\kappa}(p_\gamma)-\sum_\gamma p_\gamma^{-\kappa} \geq 0
\end{equation}
that allows us to obtain the following condition of stability:
\begin{equation}
\frac{-\sum_\gamma \ln_{\kappa}(p_\gamma)}{\sum_\gamma p_\gamma^{-\kappa}} \geq \frac{1}{1+\kappa}
\end{equation}
\begin{flushright}
$\Box$
\end{flushright}

\section{Comments and conclusions}
As Berkovitz, Frigg and Kronz pointed out (Berkovitz \textit{et al.} 2006): `The relevance of ergodic theory for the effective modelling of physical systems has come into serious question due to KAM theory. Kolmogorov (1954) formulated the theorem, which concerns Hamiltonian systems and perturbation theory (\dots) The theorem basically says that if a small perturbation is added to an integrable system with two physical degrees of freedom, then the tori with a sufficiently irrational winding number survive the perturbation. (\dots) Markus and Meyer (1974) developed some important consequences of the KAM theorem (\dots) The Markus-Meyer theorem is often informally characterized as the claim that generic Hamiltonian dynamical systems are not ergodic. This is a striking claim and its effect appears to undermine the relevancy of ergodic theory immediately. But this effect is mitigated as soon as the theorem is stated in full: the set of Hamiltonian flows that are ergodic on the energy hypersurfaces associated with each element of a dense set of energy values is of first category in the set of infinitely differentiable Hamiltonian flows. In short, the class of ergodic Hamiltonians is of first category in the set of generic Hamiltonians. A set $P$ is of first category in set $Q$, if $P$ can be represented as a countable union of nowhere dense sets in $P$. All other sets are of second category. Loosely speaking, a set of first category is the \textit{topological counterpart} to a set of measure zero in measure theory (\dots) The contrast class is \textit{second category}, which is (\dots) is the \textit{topological analogue} of a set of non-zero measure (\dots) The key phrase that has the mitigating effect is \textit{infinitely differentiable}. This restriction is substantial. It actually rules out a large class of Hamiltonian systems that are of physical interest particularly for classical statistical mechanics (\dots) it is worth contrasting the Markus-Meyer theorem with a theorem (the first) of Oxtoby and Ulam (1941): in the set of measure-preserving generalized dynamical flows, the set of dynamical flows that are not ergodic is of first category. The class of Hamiltonian flows of physical interest is substantially broader than the class considered in the Markus-Meyer theorem, but substantially narrower than that considered in the Oxtoby-Ulam theorem.
(\dots) Boltzmann (1871) studied hard ball systems in developing a mathematical foundation for statistical mechanics. He conjectured that such systems are ergodic when the number $N$ of hard balls is sufficiently large. Whether this is the case is still an open question. However, it turns out that hard ball systems are ergodic for some small $N$. Sinai (1970) provided the first rigorous demonstration that a hard ball system is ergodic. (\dots) the \textit{Boltzmann-Sinai ergodic hypothesis} (Sinai, 1963) (\dots) says that a system of $N$ hard balls on $T^2$ or $T^3$ is ergodic for any $N\geq 2$, and it was inspired by Krylov's observation, made in 1942 that hard ball systems exhibit an instability that is similar to geodesic flows on hyperbolic surfaces, known at the time to be ergodic.'.

Ergodicity is the mechanism that allows us to pass from a mechanical description of the system to a statistical one. A fundamental question is what happens if the system is not ergodic. If the system is not ergodic, we can still have a relatively simple statistical description of it if the trajectory is a fractal curve embedded in the smooth energy surface of $\Omega$ cells. If the reduced box-counting dimension of such a curve is $0 \leq d \leq 1$ then the available phase space is a fractal object embedded in the Boltzmann $6N-$dimensional smooth phase space: when $d = 1$ we regain completely the ergodicity (Garc\'{\i}a-Morales and Pellicer 2006). Starting from the analysis of the entropy and information on fractal space Naschie (Naschie 1992) showed a connection between the information and the topological dimension of some fractal sets. Moreover, he showed that the dynamics of this kind of phase space is quasi-ergodic for a dimension $N\geq 4$ of the number of particles in the fractal phase space of topological dimension $d_f$ (Wang 2004).

The unifying principle of the maximum entropy generation allows to build a generalized thermodynamic formalism analogously to that of Hamiltonian mechanics. This is clearly seen in the definition of the entropy (\ref{Lagrangian}) and (\ref{irrentrdef}). Many experiments in various fields of nuclear and condensed matter physics suggest the inadequacy of the Boltzmann-Gibbs statistics and the need for
the introduction of new statistics (Kaniadakis and Lissia 2003). In particular a large class of phenomena show power-law distributions with asymptotic long tails. Typically, these phenomena arise in presence of long-range forces, long-range memory, in chaotic systems, in opens systems or when dynamics evolves in a non-euclidean multi-fractal space-time. In the last decade many authors pursued new statistical mechanics theories that mimic the structure of the Boltzmann-Gibbs theory, while capturing the emerging
experimental anomalous behaviors (Kaniadakis and Lissia 2003). In the last ten years, an intense activity has been modifying and improving our understanding of contemporary Statistical Mechanics and Thermodynamics. 

The aim of this paper was to find an analytical relation on probability to obtain a condition on stability of steady states of open systems. This is the relation (\ref{pr}). It represents the condition of quasi-ergodicity for the open systems and it is a consequence of the principle of maximum entropy generation.


\end{document}